\documentclass[12pt]{iopart}
\usepackage{iopams}
\usepackage{bm}
\usepackage{dsfont}
\usepackage{graphicx}
\usepackage{setstack}
\usepackage[colorlinks=true,breaklinks=false,allcolors=blue]{hyperref}
\usepackage{doi}
\usepackage{url}

\newcommand\RR{{\mathds{R}}}
\newcommand\CC{{\mathds{C}}}
\newcommand\one{{\mathds{1}}}

\begin{document}

\title{Relaxation timescales and decay of correlations in a long-range interacting quantum simulator}
\author{Mauritz van den Worm$^1$, Brian C Sawyer$^2$, John J Bollinger$^2$ \& Michael Kastner$^{1,3}$}
\address{$^{1}$ National Institute for Theoretical Physics (NITheP), Stellenbosch 7600, South Africa}
\address{$^{2}$ US National Institute of Standards and Technology, Boulder, Colorado 80305, USA}
\address{$^{3}$ Institute of Theoretical Physics, University of Stellenbosch, Stellenbosch 7600, South Africa}
\ead{kastner@sun.ac.za}

\begin{abstract}
We study the time evolution of correlation functions in long-range interacting quantum Ising models. For a large class of initial conditions, exact analytic results are obtained in arbitrary lattice dimension, both for ferromagnetic and antiferromagnetic coupling, and hence also in the presence of geometric frustration. In contrast to the nearest-neighbour case, we find that correlations decay like stretched or compressed exponentials in time. Provided the long-range character of the interactions is sufficiently strong, pronounced prethermalization plateaus are observed and relaxation timescales are widely separated. Specializing to a triangular lattice in two spatial dimensions, we propose to utilize these results for benchmarking of a recently developed ion-trap based quantum simulator.
\end{abstract}


\maketitle
\tableofcontents

\section{Introduction}
When a physical system is coupled to a heat bath, one expects to observe thermalization to an equilibrium state whose temperature is determined by the bath properties. For an isolated many-body system, i.e.\ in the absence of a heat bath, the situation is less clear, although some kind of relaxation to equilibrium may be expected for sufficiently large generic systems and suitable observables. Recent progress in experiments with cold atoms and ions \cite{Kinoshita_etal06,Hofferberth_etal07,Trotzky_etal12} has stimulated intense theoretical interest in equilibration and thermalization behaviour of isolated many-body quantum systems \cite{CazalillaRigol10,Polkovnikov_etal11}. General mechanisms leading to thermalization have been proposed \cite{Deutsch91,Srednicki94,Rigol_etal08}, rigorous proofs of equilibration have been obtained for generic Hamiltonians \cite{Reimann08,Linden_etal09,Goldstein_etal10,Reimann10,ShortFarrelly12,ReimannKastner12}, and analytic as well as numeric model studies have been reported (see \cite{Polkovnikov_etal11} for a list of references). Much less is known about the timescales on which relaxation to equilibrium takes place. In this paper we study the time evolution of correlation functions in isolated long-range interacting quantum Ising systems on arbitrary lattices. We derive exact analytic results that further our understanding of different relaxation timescales and general non-equilibrium properties of long-range interacting systems. In particular, we find that for sufficiently long-range interactions, a second relaxation timescale occurs, widely separated from and substantially slower than the timescale on which single spin observables approach equilibrium. As a consequence, pronounced prethermalization plateaus are observed.

Apart from broadening the theoretical understanding, our results may also prove beneficial for the interpretation of data from cold atom or ion experiments where long-range Ising interactions can occur. For example, our results can be used to benchmark a recently developed ion-trap based quantum simulator consisting of several hundred beryllium ions stored in a Penning trap and confined to a single plane \cite{Britton_etal12}. Due to their mutual electrostatic repulsion, the ions arrange into a two-dimensional Coulomb crystal on a triangular lattice (see figure 1). The valence electron spins of the $^9$Be$^+$ ions are the relevant degrees of freedom  used for quantum simulation. Effective interactions between these spins are induced by transverse motional modes of the Coulomb crystal. Under the assumption of small, coherent ion displacements, it was shown in \cite{MintertWunderlich01,PorrasCirac04} that the resulting interactions are described by the Ising Hamiltonian
\begin{equation}\label{e:H1}
H=-\sum_{i<j}J_{i,j}\sigma_i^z\sigma_j^z-\bm B_\mu\cdot\sum_{i}\bm\sigma_i.
\end{equation}
Here $i$ and $j$ label the $N$ ions on the triangular lattice, $\bm\sigma_i=(\sigma_i^x,\sigma_i^y,\sigma_i^z)$ denotes the vector of Pauli matrices for ion $i$, the $J_{i,j}$ are coupling coefficients, and $\bm B_\mu$ is an effective magnetic field generated by externally applied microwave radiation. Unlike in the conventional Ising model \cite{Ising25}, spin--spin coupling is not restricted to nearest neighbours on the lattice, but extends over all pairs of ions. The coupling coefficients $J_{i,j}$ can be expressed in terms of the transverse phonon eigenfunctions of the lattice and a few other experimental parameters (equation (4) of \cite{Kim_etal09}). A numerical evaluation of that expression for the given lattice geometry shows that the approximation $J_{i,j}\propto D_{i,j}^{-\alpha}$ holds to a very good degree \cite{Britton_etal12}. Here $D_{i,j}$ denotes the Euclidean distance between sites $i$ and $j$ on the lattice, and $\alpha$ is an exponent that can in principle be tuned within the range $0\leqslant\alpha\leqslant3$ by varying the difference in frequency of two off-resonant lasers used in the experiment. The absolute values and even the signs of the coefficients $J_{i,j}$ can also be tuned, allowing for the investigation of ferromagnetic as well as anti-ferromagnetic couplings. In the latter case, due to geometric frustration on the triangular lattice, spin liquids and other exotic quantum phases may possibly occur. 
The results can also be applied to the linear radio-frequency ion trap quantum simulators \cite{Friedenauer_etal08,Islam_etal11,Lanyon_etal11} and, along similar lines, benchmarking of quantum magnets emulated by means of ultracold molecules is possible \cite{Hazzard_etal13}.

\section{Equilibration and thermalization of isolated quantum systems.}
\label{s:equilibration}
Equilibration and thermalization are two related, but not equivalent, notions concerning the long-time evolution of a dynamical system. Quoting Linden {\em et al.}\ \cite{Linden_etal09}, `a system equilibrates if its state evolves toward some particular state [\dots] and remains in that state (or close to it) for almost all times'. These authors then continue their description of equilibration by pointing out that the corresponding equilibrium state need not necessarily be a Gibbs state or any other special state, and it even may depend on the initial state of the system. In contrast, thermalization is a stronger notion. It requires that the equilibrium state attained does not depend on the details of the initial state, but at most on a few relevant parameters (like the total energy or the bath temperature), and that the equilibrium state is a Gibbs state (microcanonical, canonical, or possibly a generalization thereof). The calculations reported in the present paper are valid only for a certain class of initial states as specified in \sref{s:corr}, and we are therefore in no position to make claims about initial state independence. For this reason, we will speak only about equilibration in the following. 

It is important to note that the definition of equilibration does not require the time-evolved state (or density operator) to converge to the equilibrium state in the long-time limit: It is perfectly admissible to have fluctuations around the equilibrium state that do not fade away in the long-time limit, as long as such fluctuations are either sufficiently small, or large but very rare. In any quantum system on a finite-dimensional Hilbert space, time evolution is periodic or quasi-periodic. As a consequence, the aforementioned large but rare fluctuations will inevitably occur in the form of recurrences (or Loschmidt echos). We will observe such quasi-periodic behaviour for finite-system correlation functions of the long-range Ising model in \sref{s:corr}.

From a technical point of view, there are (at least) two different ways in which equilibration in the presence of fluctuations can be studied. One possible approach is probabilistic and has become known under the name of {\em typicality}\/ \cite{Goldstein_etal06,Reimann08,Linden_etal09,Goldstein_etal10,Reimann10,Jacobson_etal11,ShortFarrelly12,ReimannKastner12}. This approach amounts to studying the probability in time that the time-evolved state differs (in operator norm) from some equilibrium state by more than some small $\epsilon>0$. If this probability is less than some small $\delta(\epsilon)>0$, the system is considered as equilibrated. A second approach of dealing with fluctuations consists in taking a suitable infinite-system limit, thereby pushing the recurrence times to infinity and simultaneously making the amplitude of small fluctuations vanishingly small. We follow this second strategy in the present work.

Finally, a comment is in order on the dichotomy of closed versus open systems. We consider in the following a closed (isolated) spin system, in the sense that no thermal bath or other reservoir is coupled to the system. However, for such a closed system, we do not study the relaxation to equilibrium in the above described sense of density operators being close to the equilibrium state. Instead, we restrict our attention to the long-time behaviour of spin--spin correlation functions between lattice sites $i$ and $j$. This can be viewed as studying relaxation to equilibrium in a closed system, but only for a restricted class of observables. Alternatively, since such correlation functions are fully determined by the reduced density operator of sites $i$ and $j$, we can consider these two sites as an open system, coupled to a bath consisting of all the spins on the remaining lattice sites. We will take advantage of this open-system point of view when investigating dephasing and purities in \sref{s:prethermalization}.

\section{Long-range quantum Ising model.}
For the theoretical analysis of relaxation times and the decay of correlations, we study a model which is more general than the Hamiltonian \eref{e:H1} in some respects, and more restricted in others: We generalism to lattices of arbitrary spatial dimension $d$ and arbitrary lattice structure, but come back to the two-dimensional triangular lattice later in this paper. For our analytic calculations, the effective magnetic field $\bm B_\mu$ in \eref{e:H1} is required to point in the $z$-direction, yielding a Hamiltonian of the form
\begin{equation}\label{e:H2}
H_\ell=-\sum_{i<j}J_{i,j}\sigma_i^z\sigma_j^z-B\sum_{i}\sigma_i^z,
\end{equation}
where the index $\ell$ indicates the presence of a longitudinal field $B$. At first sight it may look as if the resulting problem is purely classical, as all terms in $H_\ell$ commute with each other, and equilibrium properties such as the partition function are indeed identical to those of the corresponding classical Ising model. For non-equilibrium calculations, however, quantumness enters through observables and/or initial density operators that do not commute with the Hamiltonian \eref{e:H2}, and indeed one can prove that entanglement and other genuine quantum properties are generated under the time evolution of \eref{e:H2}. Observables consisting only of $\sigma_i^z$ operators commute with the Hamiltonian and therefore show no relaxation behaviour, but other observables do, as we will confirm in the following. In spite of its simplicity, the model \eref{e:H2} exhibits non-trivial and remarkably rich non-equilibrium behaviour.

In equilibrium, the long-range interacting Ising model \eref{e:H2} with power law decaying interaction strength $J_{i,j}\propto D_{i,j}^{-\alpha}$ is known to undergo a transition from a ferromagnetic phase at low temperature to a paramagnetic phase at high temperature. In spatial dimension $d\geqslant2$ such a transition occurs for all nonnegative values of the exponent $\alpha$, whereas for $d=1$ the transition is present only for $\alpha\leqslant2$ \cite{Dyson69a}. Remarkably and surprisingly, the long-time dynamics investigated in this paper turns out to be independent on whether the corresponding energies are situated in the low- or high-energy regime.

\section{Time evolution of correlation functions.}
\label{s:corr}
To study the relaxation to equilibrium in this model, our aim is to compute the time evolution of spin--spin correlation functions
\begin{equation}\label{e:spinspin1}
\left\langle\sigma_i^x\sigma_j^x\right\rangle(t)=\Tr \left(\rme^{\rmi H_\ell t} \sigma_i^x\sigma_j^x \rme^{-\rmi H_\ell t}\rho_0\right).
\end{equation}
Here and in the following, we set $\hbar=1$ for convenience. For analytical calculations, a significant simplification is achieved by restricting the initial states to density operators $\rho_0$ that are diagonal matrices in the $\sigma^x$ tensor-product eigenbasis,
\begin{equation}\label{e:rho0}
\rho_0=\frac{1}{2^N}\biggl(\one+\sum_{i}\sigma_i^x\biggl(s_i^x + \sum_{j>i}\sigma_j^x\biggl(s_{ij}^{xx} + \sum_{k>j}\sigma_k^x\biggl(s_{ijk}^{xxx} + \sum_{l>k}\cdots\biggr)\biggr)\biggr)\biggr),
\end{equation}
where $\one$ denotes the identity operator on the tensor product Hilbert space $\mathcal{H}=\CC^2\otimes\cdots\otimes\CC^2$. The indices $i$, $j$, $k$, $l$ in \eref{e:rho0} are summed over the lattice sites. The set of all $\rho_0$ as defined in \eref{e:rho0} is is a very large class of initial states, parametrized by $2^N-1$ real continuous parameters $s_i^x=\left\langle\sigma_i^x\right\rangle(0)$, $s_{ij}^{xx}=\left\langle\sigma_i^x\sigma_j^x\right\rangle(0),\dots\;$ The reader may wish to compare the size of this class to the much studied quantum quenches, a rather restrictive class of initial states parametrized by only two parameters (namely the quench parameter before and after the quench). We could, as a matter of fact, extend the class of initial states \eref{e:rho0} even further, allowing also $\sigma^y$-type contributions to $\rho_0$. This extension essentially leaves the results of the present paper unaltered, but the presentation becomes somewhat more involved. We decided to discuss this generalization in a future paper devoted to the more technical aspects of long-range quantum spin models \cite{Rey+}.

The idea of considering initial states of the form \eref{e:rho0} has been used in \cite{Emch66,Kastner11,Kastner12} for the computation of expectation values of one-spin observables. In this paper, these restrictions are relaxed and the exact analytic calculations are extended to correlation functions for arbitrary finite system sizes. Details of the calculation are reported in \ref{A}. For vanishing longitudinal magnetic field $B=0$, the final results are
\numparts
\begin{eqnarray}
\left\langle\sigma_i^x\sigma_j^y\right\rangle\!(t)&=&\left\langle\sigma_i^x\sigma_j^z\right\rangle\!(t)=\left\langle\sigma_i^z\sigma_j^z\right\rangle\!(t)=0,\label{e:ss1}\\
\left\langle\sigma_i^y\sigma_j^z\right\rangle\!(t)&=&\left\langle\sigma_i^x\right\rangle\!(0)\sin\left(2tJ_{i,j}\right)\prod_{k\neq i,j}\cos\left(2tJ_{i,k}\right),\label{e:ss2}\\
\left\langle\sigma_i^x\sigma_j^x\right\rangle\!(t)&=&P_{i,j}^- + P_{i,j}^+,\quad\left\langle\sigma_i^y\sigma_j^y\right\rangle\!(t)=P_{i,j}^- - P_{i,j}^+,\label{e:ss3}
\end{eqnarray}
with
\begin{equation}
P_{i,j}^\pm=\frac{1}{2}\left\langle\sigma_i^x\sigma_j^x\right\rangle\!(0)\prod_{k\neq i,j}\cos\left[2\left(J_{k,i}\pm J_{k,j}\right)t\right].\label{e:Ppm}
\end{equation}
\endnumparts
These expressions are valid for arbitrary coupling constants $J_{i,j}$, and therefore apply in arbitrary spatial dimension and on arbitrary lattices. Besides an overall factor $\langle\sigma_i^x\rangle(0)$ or $\langle\sigma_i^x\sigma_j^x\rangle(0)$, equations \eref{e:ss1}--\eref{e:Ppm} do not depend on the particular choice of the initial state, but apply to all $\rho_0$ from the large class of initial states as given in \eref{e:rho0}. Moreover, the equations can easily be evaluated on a personal computer for millions of spins. The presence of a longitudinal magnetic field $B\neq0$ modifies equations \eref{e:ss1}--\eref{e:Ppm} by imposing an additional spin precession at an angular frequency proportional to $B$ (see \ref{A}). For the relaxation to equilibrium we are interested in, such an oscillatory spin precession is irrelevant and we will therefore restrict the discussion to the $B=0$ case in the following. A generalization of the calculation in \ref{A} to arbitrary $n$-spin correlation functions is also feasible by similar methods, but is not reported here. Since any operator on $\CC^{2^N}$ can be expanded in terms of products of Pauli operators, such results for correlation functions of arbitrary order permit, in principle, the computation of the time evolution of expectation values of arbitrary operators. This does not add much to the discussion of the physical phenomena we are interested in here, and we will therefore restrict the discussion to two-spin correlations.

\begin{figure}
{\center
\includegraphics[width=0.284\linewidth]{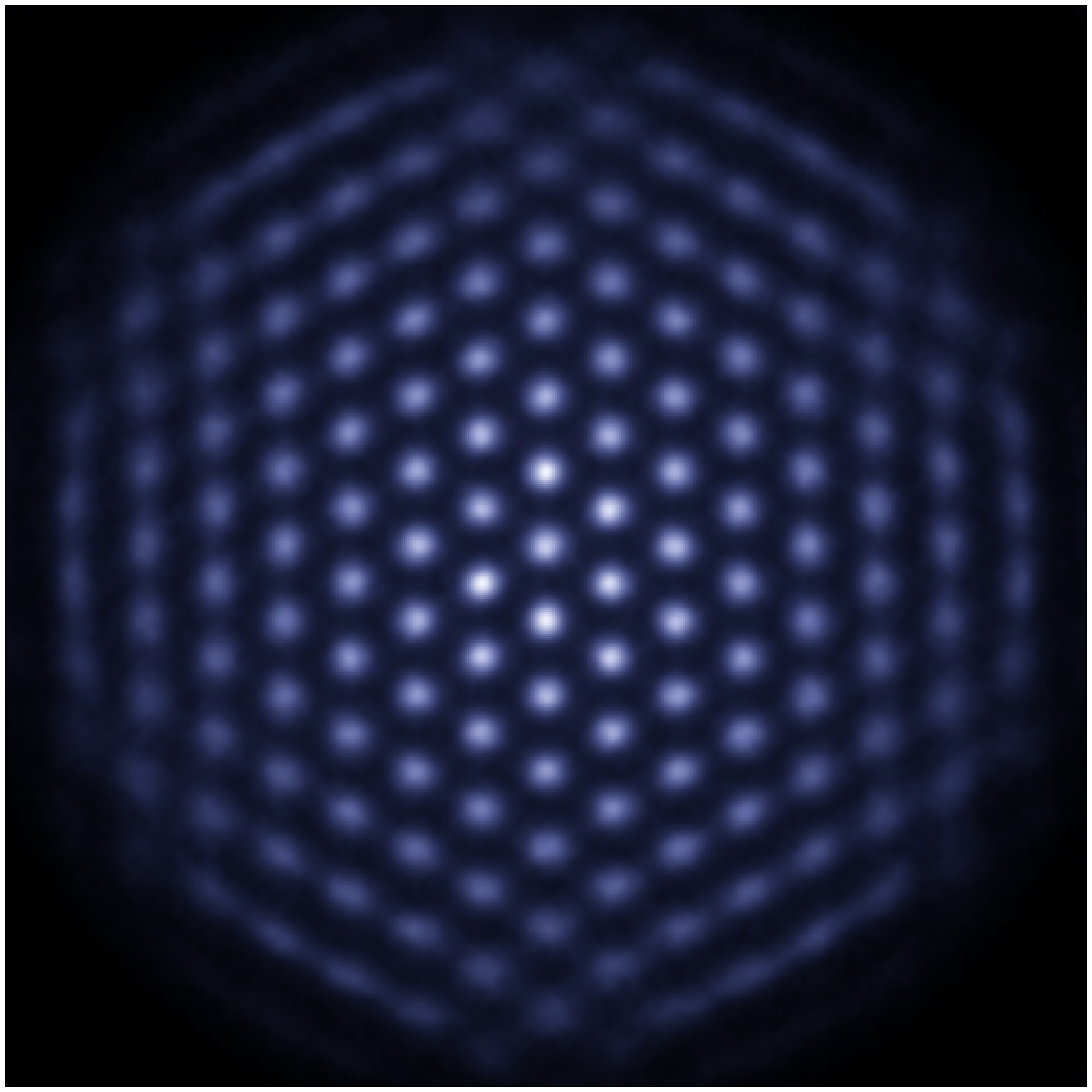}
\hfill
\includegraphics[width=0.32\linewidth]{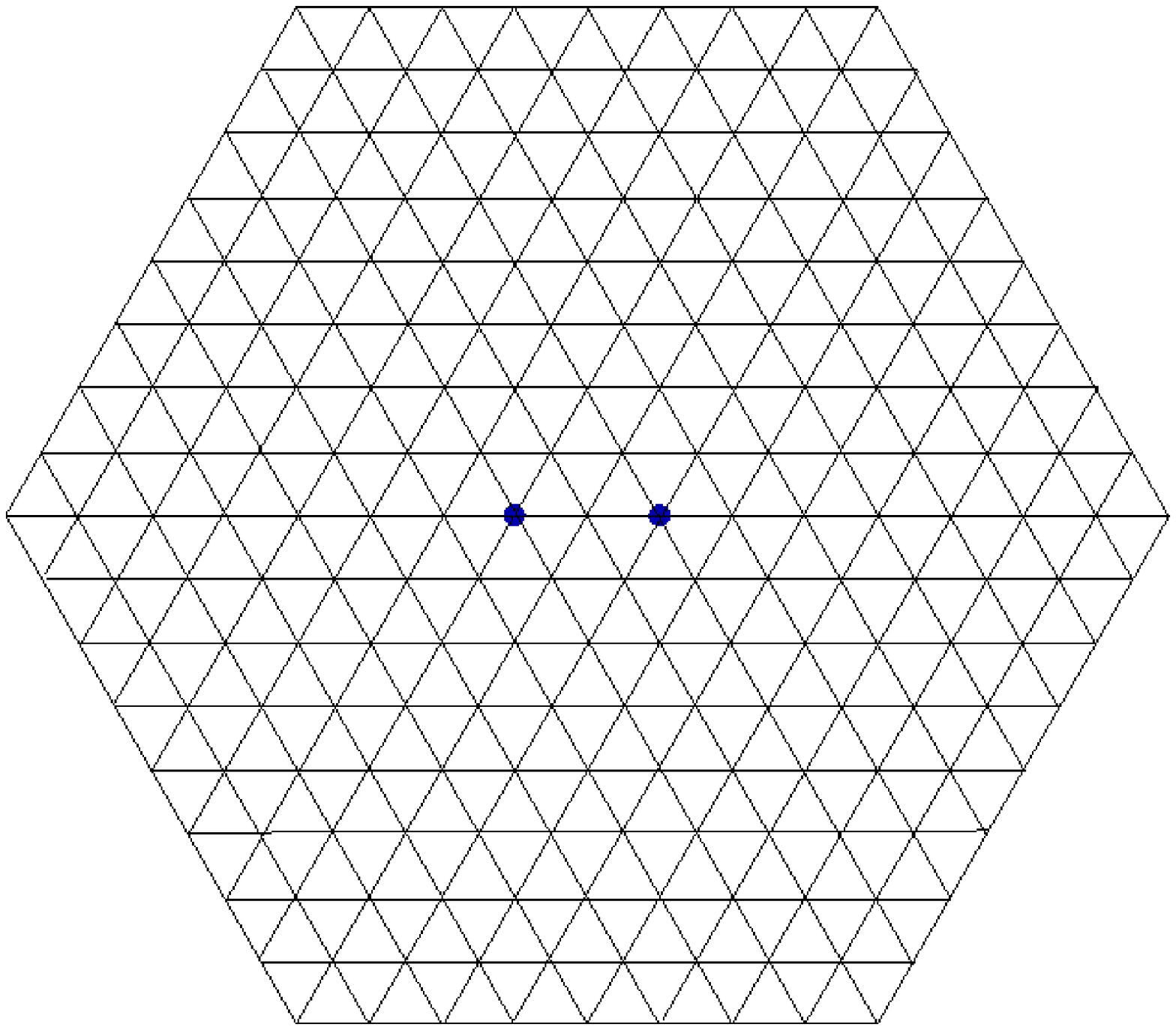}
\hfill
\includegraphics[width=0.32\linewidth]{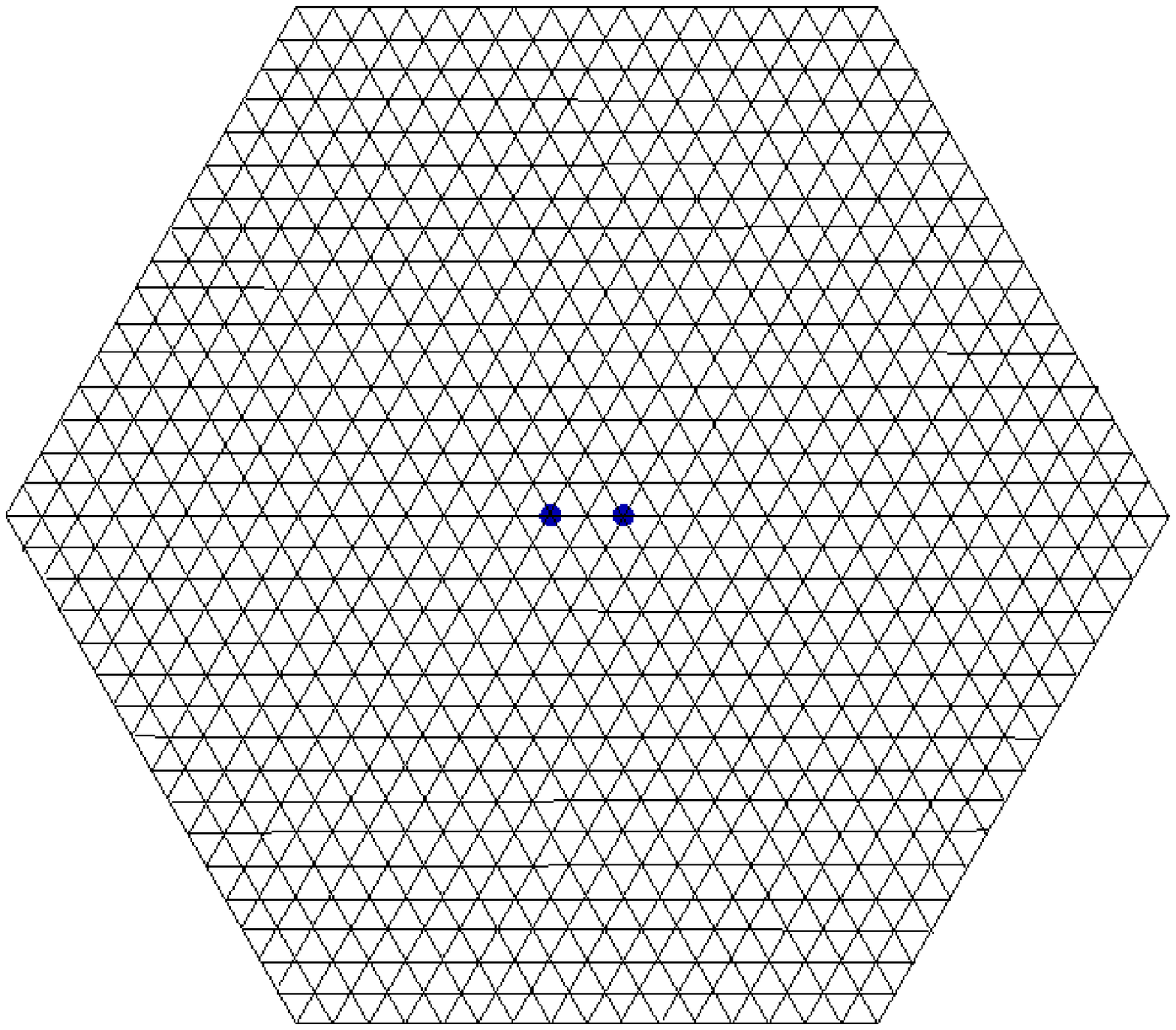}
}
\caption{\label{f:lattice}%
Left: a top-view resonance fluorescence image showing the centre region of an ion crystal captured in the ions' rest frame. Fluorescence is an indication of a valence electron being in the spin-up state. Here, all ions are in the spin-up state. The lattice constant is approximately 20 $\mu$m. Centre and right: triangular lattices on hexagonal patches of side lengths $L=8$ and $16$.
}%
\end{figure}

As an example, the time evolution \eref{e:ss1}--\eref{e:Ppm} of spin--spin correlation functions is illustrated for system parameters similar to those of the ion trap experiments of Britton {\em et al.}\ \cite{Britton_etal12}. We consider hexagonal patches of triangular lattices (\fref{f:lattice}), and couplings $J_{i,j}=JD_{i,j}^{-\alpha}$ proportional to the $\alpha$th power of the inverse Euclidean distance of sites $i$ and $j$ on that lattice, with $J\in\RR$ being a coupling constant. We have also performed numerical calculations for other two-dimensional lattice structures and geometries, and the behaviour we found is qualitatively the same and also quantitatively very similar. 

\begin{figure}
{\center
\includegraphics[width=0.48\linewidth]{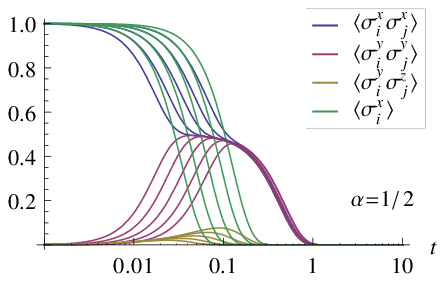}
\hfill
\includegraphics[width=0.48\linewidth]{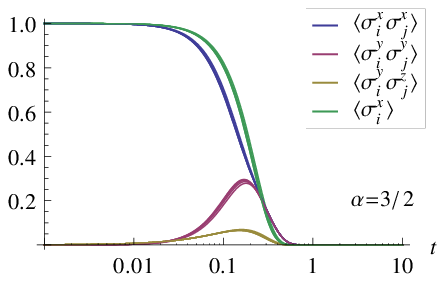}
}
\caption{\label{f:evolution}%
Time evolution of the normalized expectation values $\langle\sigma_i^x\sigma_j^x\rangle(t)/\langle\sigma_i^x\sigma_j^x\rangle(0)$, $\langle\sigma_i^y\sigma_j^y\rangle(t)/\langle\sigma_i^x\sigma_j^x\rangle(0)$, $\langle\sigma_i^y\sigma_j^z\rangle(t)/\langle\sigma_i^x\rangle(0)$, $\langle\sigma_i^x\rangle(t)/\langle\sigma_i^x\rangle(0)$, as given by \eref{e:ss1}--\eref{e:Ppm}, for the long-range Ising model on a triangular lattice. Normalized in this way, the results shown are valid for all initial conditions of the form \eref{e:rho0}. The lattice sites $i$ and $j$ are chosen one lattice site to the right, respectively left, of the centre of the hexagonal patch, as indicated by the blue dots in \fref{f:lattice} (centre and right). The various curves of the same colour in each plot correspond to side lengths $L=4$, $8$, $16$, $32$, and $64$ (from right to left) of the hexagonal patches. The left panel is for power law interactions with exponent $\alpha=1/2$, but results are qualitatively similar for all $\alpha$ between zero and $d/2$. The right panel is for $\alpha=3/2$, with qualitatively similar results for all $\alpha>d/2$. The unit of time is $1/J$.
}%
\end{figure}

The plots in \fref{f:evolution} show the normalized expectation values $\langle\sigma_i^x\sigma_j^x\rangle(t)/\langle\sigma_i^x\sigma_j^x\rangle(0)$, $\langle\sigma_i^y\sigma_j^y\rangle(t)/\langle\sigma_i^x\sigma_j^x\rangle(0)$, $\langle\sigma_i^y\sigma_j^z\rangle(t)/\langle\sigma_i^x\rangle(0)$, $\langle\sigma_i^x\rangle(t)/\langle\sigma_i^x\rangle(0)$, as given by \eref{e:ss1}--\eref{e:Ppm}. Normalized in this way, the plots are valid and numerically exact for all initial conditions of the form \eref{e:rho0}, and the same is true for figures \ref{f:evolution2}, \ref{f:purity}, and \ref{f:bounds}. The plotted curves suggest that correlations decay to their microcanonical equilibrium values $\langle\sigma_i^a\sigma_j^b\rangle_{\rm{mc}}=0$ where $a,b\in\{x,y,z\}$, but this decay is only apparent: Since the expressions in \eref{e:ss2}--\eref{e:Ppm} consist of products of $N$ trigonometric functions, the evolution in time of spin--spin correlations is quasi-periodic for any finite number $N$ of spins. This is exactly the situation described in \sref{s:equilibration} for general quantum systems on a finite-dimensional Hilbert space. Accordingly, recurrences will occur and repeatedly bring the system arbitrarily close to its initial state. These recurrences occur on a timescale that is exponentially large in $N$ and, already for moderate system sizes, they do not show up on the timescales plotted in \fref{f:evolution}. (For example, for a lattice with side length $L=4$ and $\alpha=3/2$, the first significant recurrence occurs at a time three orders of magnitude larger than the relaxation time). In the large-$N$ limit, this recurrence time is expected to be pushed to infinity. We will confirm this expectation in \sref{s:bound} by proving, in the thermodynamic limit, an upper bound on spin--spin correlation functions which converges to zero for large times $t$.

\section{Upper bound on correlations in the thermodynamic limit.}
\label{s:bound}
With \eref{e:ss2}--\eref{e:Ppm} as a starting point, it is possible to derive upper bounds on the time evolution of spin--spin correlations in the thermodynamic limit for different lattice structures (see \ref{B} for the derivation in the case of a triangular lattice). 
All these bounds are of the form
\numparts
\begin{equation}
\left\langle\sigma_i^x\sigma_j^x\right\rangle\!(t),\;\left\langle\sigma_i^y\sigma_j^y\right\rangle\!(t)\leqslant\bar{P}_{i,j}^- + \bar{P}_{i,j}^+\label{e:boundxxyy}
\end{equation}
with
\begin{equation}
\bar{P}_{i,j}^\pm=\frac{1}{2}\left\langle\sigma_i^x\sigma_j^x\right\rangle\!(0)\exp\bigl[-C_{i,j}^\pm(\alpha)N^{q^\pm(\alpha)}t^{p^\pm(\alpha)}\bigr]\label{e:Pbarpm}
\end{equation}
and
\begin{eqnarray}
q^+(\alpha)&=&\max\{0,1-2\alpha/d\},\\
q^-(\alpha)&=&\max\{0,1-2(1+\alpha)/d\},\\
p^+(\alpha)&=&\min\{2,d/\alpha\},\\
p^-(\alpha)&=&\min\{2,d/(1+\alpha)\}.
\end{eqnarray}
\endnumparts
The constants $C_{i,j}^\pm$ in \eref{e:Pbarpm} depend on the lattice structure, the exponent $\alpha$, and the position of $i$ and $j$ on the lattice. Other 2-spin correlation functions have similar properties, and details of the derivation are  provided in \ref{B}. All the bounds in the above equations decay to zero for large $t$. As can be read off from these equations, the slowest decay is always due to the $\bar{P}_{i,j}^-$-term. For $d\leqslant2$, relaxation to zero is therefore governed by a term of the form $\exp\!\left[-C_{i,j}^-(\alpha)t^{d/(1+\alpha)}\right]$, implying a compressed exponential decay for $\alpha<d-1$, and stretched exponential decay otherwise.

\begin{figure}
{\center
\includegraphics[width=0.5\linewidth]{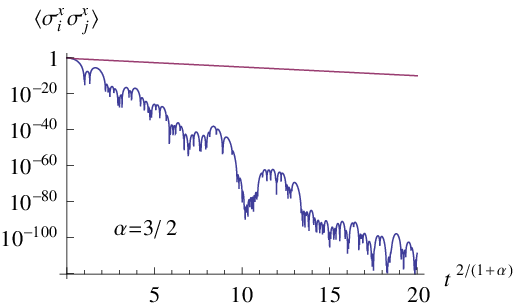}
\hfill
\includegraphics[width=0.43\linewidth]{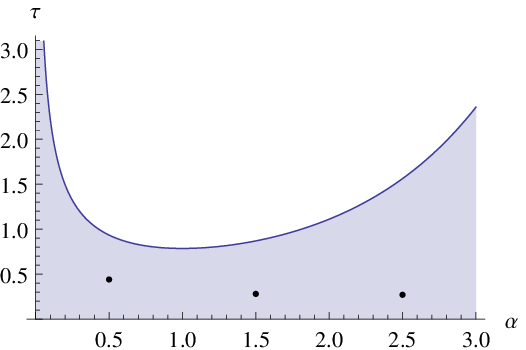}
}
\caption{\label{f:evolution2}%
Left: upper bound \eref{e:boundxxyy} (straight red line) compared to the exact time evolution \eref{e:ss3} (wiggly blue line) of the spin--spin correlator $\langle\sigma_i^x\sigma_j^x\rangle(t)/\langle\sigma_i^x\sigma_j^x\rangle(0)$ for a triangular lattice with $\alpha=3/2$ and $L=32$, plotted {\em vs.}\ the rescaled time $t^{2/(1+\alpha)}$. The linearly decaying trend in the plot, superimposed by fluctuations, confirms the asymptotic behaviour over a range of more than hundred orders of magnitude. Right: plot of the upper bound \eref{e:taubound} on the relaxation time {\em vs.}\ the exponent $\alpha$. According to this bound, for any given value of $\alpha$ the actual value of $\tau$ must lie in the shaded area. For comparison, finite-system relaxation times as read off from \fref{f:evolution} for $\alpha=1/2$ and $3/2$ (and from a similar plot for $5/2$) are shown as black dots.
}%
\end{figure}

For the example of a triangular lattice in two dimensions, we obtain
\numparts
\begin{eqnarray}
C_{i,j}^+(\alpha) &=& 
\cases{\displaystyle\frac{(8J)^2\,3^{\alpha-1}}{\pi^2(1-\alpha)} & for $\alpha<1$,\\
\displaystyle\frac{1}{\alpha-1}\left(\frac{8J}{\pi}\right)^{2/\alpha} & for $\alpha>1$,}\\[1mm]
C_{i,j}^-(\alpha) &=& \frac{D_{i,j}^2}{4\alpha}\left(\frac{4\alpha J}{\pi}\right)^{2/(1+\alpha)}.\label{e:C-}
\end{eqnarray}
\endnumparts
With $i,j$ as in \fref{f:lattice} we have $D_{i,j}=2$, and setting $\alpha=3/2$ we obtain $C_{i,j}^+\approx11.04$ and $C_{i,j}^-\approx1.119$. We inserted these values into the bound \eref{e:boundxxyy} and compared the outcome to exact finite-system results. \Fref{f:evolution2} (left) shows the correlation function $\left\langle\sigma_i^x\sigma_j^x\right\rangle$ and its upper bound \eref{e:boundxxyy} {\em vs.}\ the rescaled time $t^{2/(1+\alpha)}$. The rescaling of time is chosen such that the upper bound \eref{e:boundxxyy} is a linearly decaying function. The plot reveals that, over a range of more than hundred orders of magnitude, the exact finite-system result for $\left\langle\sigma_i^x\sigma_j^x\right\rangle$ likewise decays with a linear trend, superimposed by fluctuations. This confirms that the  upper bound \eref{e:boundxxyy} indeed correctly captures the stretched or compressed exponential decay of the correlation function, albeit with prefactors $C_{i,j}^\pm$ that overestimate the actual behaviour.

The stretched or compressed exponential relaxation of the Ising model with power law interactions discussed above is different from the exponential decay of correlations known to occur in the nearest-neighbour Ising model \cite{CalabreseEsslerFagotti11}. In the limit $\alpha\to\infty$ of the power law interactions one would usually expect to approach the nearest-neighbour-interacting model and recover exponentially decaying correlations, but this does not seem to be the case. This puzzle is resolved by observing that, in the limit $\alpha\to\infty$, the coefficient $C_{i,j}^-$ in \eref{e:C-} also diverges, implying that the bound \eref{e:boundxxyy} is not meaningful in this limit.

On the basis of the bound \eref{e:Pbarpm} on $\bar{P}_{i,j}^-$, one can define an upper bound on the relaxation time by requiring the exponent in \eref{e:Pbarpm} to be, say, minus unity,
\begin{equation}
C_{i,j}^-(\alpha)t^{d/(1+\alpha)}=1.
\end{equation}
Inserting \eref{e:C-} and solving for $t$, an upper bound is obtained on the relaxation time $\tau$ for a triangular lattice in the thermodynamic limit,
\begin{equation}\label{e:taubound}
\tau(\alpha)\leqslant\left(\frac{2\sqrt{\alpha}}{D_{i,j}}\right)^{1+\alpha}\frac{\pi}{4\alpha J}.
\end{equation}
In \fref{f:evolution2} (right) this bound on the relaxation time is plotted {\em vs.}\ the exponent $\alpha$. Finite-system relaxation times as read off from \fref{f:evolution} for $\alpha=1/2$ and $\alpha=3/2$ are also shown for comparison.

\section{Prethermalization.}
\label{s:prethermalization}
Apart from the overall relaxation times, there are other striking $\alpha$-dependent aspects of the relaxation dynamics: The plot in the right panel of \fref{f:evolution}, representative for exponents $\alpha\geqslant d/2$, shows a simple relaxation to equilibrium with a single relevant timescale. The plot in the left panel of \fref{f:evolution}, representative for exponents $0\leqslant\alpha\leqslant d/2$, differs considerably: In a first step, the correlation functions $\left\langle\sigma_i^x\sigma_j^x\right\rangle$ and $\left\langle\sigma_i^y\sigma_j^y\right\rangle$ display a Gaussian decay on a fast timescale $\tau_1$ to one half of the $\left\langle\sigma_i^x\sigma_j^x\right\rangle$ initial value, before finally relaxing to their vanishing equilibrium value on a much longer timescale $\tau_2$. $\tau_1$ is roughly the same as the timescale on which one-spin observables (green lines in \fref{f:evolution}) relax, while correlations are not yet equilibrated. Then the system remains {\em prethermalized} \cite{MoeckelKehrein08} for a relatively long period of time, a behaviour observed in the left panel of \fref{f:evolution} as a conspicuous plateau. Different from earlier observations of prethermalization in quantum dynamics, the ratio $\tau_1/\tau_2$ goes to zero in the large-$N$ limit, i.e.\ separation of timescales become more pronounced with increasing system size. Such a behaviour of $N$-dependent relaxation timescales had previously been observed in classical long-range interacting systems, either for mean-field-type spin models \cite{Yamaguchi_etal04}, or for self-gravitating systems in the astrophysical context \cite{Gabrielli_etal10}. Besides the large-$N$ limit, also the $\alpha\to0$ limit leads to a more pronounced separation of timescales in the long-range Ising model. In the $\alpha\to0$ limit, the slow relaxation timescale $\tau_2$ diverges, while the fast timescale $\tau_1$ remains finite for finite $N$. This fact can be explained by the presence of $N(N-1)$ additional symmetries in the $\alpha=0$ case, where all operators $\sigma_i^x\sigma_j^x+\sigma_i^y\sigma_j^y$ commute with the Hamiltonian \eref{e:H2}. Although these symmetries are absent for $\alpha\gtrsim0$, the slow timescale $\tau_2$ can be seen as a remnant of the weakly destroyed $\alpha=0$ symmetry. While two-spin correlations exhibit relaxation in a two-step process for $\alpha<d/2$, we have checked that no third timescale emerges when the calculations are generalized to three-spin correlations.

\begin{figure}\centering
\includegraphics[width=0.5\linewidth]{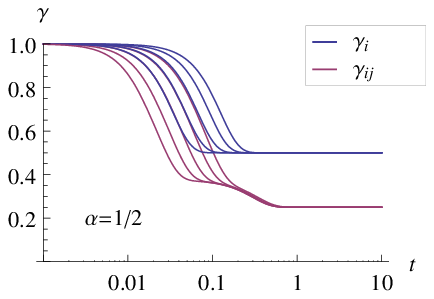}
\includegraphics[width=0.48\linewidth]{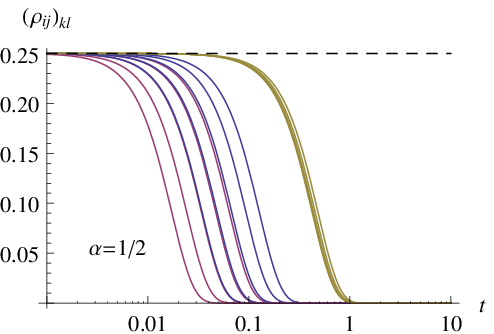}
\caption{\label{f:purity}%
The various curves of the same colour correspond to side lengths $L=4$, $8$, $16$, $32$, and $64$ (from right to left) of the hexagonal patches of triangular lattices. Left: time evolution of the one- and two-spin purities $\gamma_i$ and $\gamma_{ij}$ as defined in \eref{e:purity}. Each relaxation step in spin--spin correlations in \fref{f:evolution} is accompanied by a drop in $\gamma_{ij}$. Right: time evolution of the modulus of the matrix elements of the 2-spin reduced density matrix \eref{e:rhoij}. The black dashed line corresponds to the diagonal elements \eref{e:black}, red corresponds to \eref{e:red}, yellow to \eref{e:yellow}, and blue to \eref{e:blue}.
}%
\end{figure}

Multiple timescales of relaxation may emerge for various reasons. One possible scenario that has been observed previously, both in theory and experiment \cite{Gring_etal12}, is that dephasing is responsible for prethermalization, while a collisional mechanism causes relaxation on a slower timescale. In order to test which of these mechanism is at work in the long-range Ising model, we have computed the time evolution of the $n$-spin purity
\begin{equation}\label{e:purity}
\gamma_{i_1,\dots,i_n}(t)=\Tr\left(\rho_{i_1,\dots,i_n}^2(t)\right),
\end{equation}
where
\begin{equation}
\rho_{i_1,\dots,i_n}=\Tr_{\Lambda\setminus\{i_1,\dots,i_n\}}\rho
\end{equation}
is the $n$-spin reduced density operator, as obtained by tracing the density operator $\rho$ over all sites of the lattice $\Lambda$ except $i_1,\dots,i_n$. Knowledge of the one-spin expectation values $s_i^a=\langle\sigma_i^a\rangle$ with $a\in\{x,y,z\}$ allows for the reconstruction of
\begin{equation}
\rho_i=\frac{1}{2}\left(\!\!\begin{array}{cc}1+s_i^z & s_i^x-\rmi s_i^y\\s_i^x+\rmi s_i^y & 1-s_i^z\end{array}\!\!\right),
\end{equation}
and additional knowledge of the spin--spin correlations $s_{ij}^{ab}$ with $a\in\{x,y,z\}$ facilitates the reconstruction of 
\setlength{\arraycolsep}{1.37mm}
\begin{eqnarray}
\fl\rho_{i,j}=\frac{1}{4}\!\left(\!\!\footnotesize{\begin{array}{cccc}
1+s_i^z+s_j^z+s_{ij}^{zz} & s_j^x-\rmi s_j^y+s_{ij}^{zx}-\rmi s_{ij}^{zy} & s_i^x-\rmi s_i^y+s_{ij}^{xz}-\rmi s_{ij}^{yz} & s_{ij}^{xx}-s_{ij}^{yy}-\rmi s_{ij}^{xy}-\rmi s_{ij}^{yx}\\
& 1+s_i^z-s_j^z-s_{ij}^{zz} &  s_{ij}^{xx}+s_{ij}^{yy}+\rmi s_{ij}^{xy}-\rmi s_{ij}^{yx} & s_i^x-\rmi s_i^y-s_{ij}^{xz}+\rmi s_{ij}^{yz}\\
& & 1-s_i^z+s_j^z-s_{ij}^{zz} & s_j^x-\rmi s_j^y-s_{ij}^{zx}+\rmi s_{ij}^{zy}\\
\mbox{h.\,c.} & & & 1-s_i^z-s_j^z+s_{ij}^{zz}
\end{array}
}\!\!\right)\!.\nonumber\\\label{e:rhoij}
\end{eqnarray}
Inserting our results for the time evolution of spin expectation values of the long-range Ising model into these expressions, we readily obtain the time evolution of the one- and two-spin purities \eref{e:purity}. As can be seen in \fref{f:purity} (left), both relaxation steps of spin--spin correlations we observed in \fref{f:evolution} turn out to be associated with a drop in the purity $\gamma_{ij}$, which indicates that both relaxation steps are caused by dephasing.

To demonstrate the dephasing yet more clearly, we also studied directly the modulus of the off-diagonal elements of the 2-spin density operator $\rho_{ij}$ in the $\sigma_z$ tensor product eigenbasis. For vanishing external magnetic field $B=0$ and the class of initial conditions \eref{e:rho0}, we have $s_i^y=s_i^z=s_{ij}^{xy}=s_{ij}^{xz}=s_{ij}^{yx}=s_{ij}^{zx}=s_{ij}^{zz}=0$ for all times $t$. With this simplification and the additional symmetries $s_i^a=s_j^a$ and $s_{ij}^{ab}=s_{ij}^{ba}$, only four of the 16 matrix elements of $\rho_{ij}$ have different moduli,
\numparts
\begin{eqnarray}
\left|\left(\rho_{ij}\right)_{kk}\right| & = & 1/4 \qquad\forall k\in\{1,2,3,4\},\label{e:black}\\
\left|\left(\rho_{ij}\right)_{14}\right| & = & \left|\left(\rho_{ij}\right)_{41}\right| = \left|s_{ij}^{xx}-s_{ij}^{yy}\right|/4=P_{i,j}^+/2,\label{e:red}\\
\left|\left(\rho_{ij}\right)_{23}\right| & = & \left|\left(\rho_{ij}\right)_{32}\right| = \left|s_{ij}^{xx}+s_{ij}^{yy}\right|/4=P_{i,j}^-/2,\label{e:yellow}\\
\left|\left(\rho_{ij}\right)_{12}\right| & = & \left|\left(\rho_{ij}\right)_{13}\right| = \left|\left(\rho_{ij}\right)_{21}\right| = \left|\left(\rho_{ij}\right)_{24}\right| = \left|\left(\rho_{ij}\right)_{31}\right| = \left|\left(\rho_{ij}\right)_{34}\right|\nonumber\\
& = & \left|\left(\rho_{ij}\right)_{42}\right| = \left|\left(\rho_{ij}\right)_{43}\right| = \left|s_i^x+\rmi s_{ij}^{yz}\right|/4.\label{e:blue}
\end{eqnarray} 
\endnumparts
We have plotted these quantities in \fref{f:purity} (right) for several sizes of triangular lattices. All nondiagonal matrix elements decay to zero, but they do so on two different timescales: The matrix elements $\left(\rho_{ij}\right)_{23}$ and $\left(\rho_{ij}\right)_{32}$ (yellow in \fref{f:purity} (right)) are given by $P_{i,j}^-/2$ and hence decay on the longer of the two timescales we had previously observed. All other non-diagonal matrix elements decay on faster, $N$-dependent timescales. This demonstrates that both steps of the two-step relaxation process we observed for spin--spin correlation functions in \fref{f:evolution} are caused by dephasing. All the diagonal elements $\left(\rho_{ij}\right)_{kk}$ (black dashed line in \fref{f:purity} (right)) are constant in time, and therefore no redistribution of mode occupation numbers takes place. Such a redistribution would signal collisional relaxation due to inelastic scattering processes between these modes. Since the Ising model \eref{e:H2} has only commuting terms, no inelastic collisional mechanism is available.

\section{Application to trapped-ion quantum simulators.}
Long-range Ising interactions between effective two-level systems or spins can be implemented in crystalline arrays of trapped ions. Qualitatively, spin-dependent forces are used to modify the Coulomb interaction energy of the trapped ions and generate an effective Ising interaction. With small numbers of ions in linear radio-frequency traps, long-range Ising interactions like those in equation \eref{e:H1} are the basis of quantum gates, and have been implemented with high fidelity ($>$98\%). These techniques have been recently extended to single-plane triangular lattices of several hundred ions stored in a Penning trap. To date, the Ising interactions implemented in the Penning ion trap simulator have been benchmarked for short timescales through comparison with mean field theory \cite{Britton_etal12}, a classical limit of the Hamiltonian \eref{e:H1} where quantum fluctuations and correlations are ignored. The exact results on correlation functions reported here will enable a much higher level of benchmarking, and a determination of the timescales over which the expected emulation of quantum effects of an Ising model is indeed realized.
 
Application of the results derived here for the ion trap simulators requires initial states that are diagonal in the $\sigma^{x}$ tensor product eigenbasis, and assurance that the trapped ion simulator is an isolated system for the quantum many-body equilibration time scales $P_{i,j}^\pm$ discussed in the previous sections. Preparation of diagonal states in the $\sigma^{x}$ tensor product eigenbasis is straight forward in ion traps. Specifically, optical pumping and coherent control techniques can initialize each spin to point along the $x$-axis with high fidelity.  In the Penning ion trap simulator this initialization can be done with a fidelity of greater than 99\% \cite{Biercuk_etal09}. Equilibration time scales $P_{i,j}^\pm$ (see equation \eref{e:Ppm}) that are short compared with other relaxation time scales have been demonstrated with 10--20 ions in linear radio-frequency traps \cite{Kim_etal09,Friedenauer_etal08,Islam_etal11,Lanyon_etal11}. In the current Penning trapped-ion simulator \cite{Britton_etal12}, spontaneous emission from the off-resonant laser beams used to engineer the long-range Ising interaction sets an experimental relaxation timescale that is comparable to the quantum many-body equilibration time scales. However, with straight-forward improvements to the setup described in \cite{Britton_etal12}, the quantum many-body relaxation timescales can be short compared to the spontaneous emission timescale. Specifically, with $N=217$ ions and an optical dipole force generated from two 10 mW beams crossing with an angular separation of $35^{o}$ (see figure 1 of \cite{Britton_etal12}) and frequency difference tuned to generate an $\alpha=1/2$ power law interaction, we calculate the two relaxation timescales $P_{i,j}^\pm$ in \eref{e:Ppm} to be $\approx30$ $\mu$s and $\approx430$ $\mu$s. Here $i$ and $j$ are chosen to be two sites on opposite sides of the center site as shown in \fref{f:lattice}. The spontaneous emission time for this configuration is $(1/\Gamma)\sim 4$ ms. All of these timescales are short compared to the $T_{2}\gtrsim 50$ ms coherence time of the Be$^{+}$ valence electron spin qubits.

To compare with the exact results reported here requires an experimental mea\-sure\-ment of the spin--spin correlations after the Ising interaction has been applied for a variable period. This is readily accomplished with trapped ions by using spin-dependent resonance fluorescence. Britton {\em et al}\/ \cite{Britton_etal12} use spin-dependent resonance fluorescence to measure spin orientation in the $\sigma^{z}$ basis. With resolved imaging of the ions array (see experimental image in \fref{f:lattice}), the spin orientation of each ion can be detected and arbitrary pair correlation functions calculated. However, more simply, the global fluorescence collected from all the ions in the array can be used to measure the global spin state of the system (i.e., the total number of spins in the $\left|\uparrow\right\rangle_{z}$ state and the total number of spins in the $\left|\downarrow\right\rangle_{z}$ state). Shot-to-shot fluctuations in these measurements are sensitive to the second-order moment $\left\langle J_{z}^{2}\right\rangle $ of the total $z$-component of the spin $J_{z}\equiv\sum_{i=1}^{N}\sigma_{i}^{z}/2$. Of particular interest are measurements of the second-order moments in directions perpendicular to the mean composite spin vector. For example, with all spins initially pointing along the $x$-axis, fluctuations in measurements of $J_{z}$ after rotation about the $x$-axis by an angle $\theta$ are sensitive to
\begin{equation}
\fl\left\langle J_{z}^{2}\right\rangle _{\theta}=\sum_{i,j=1}^{N}\frac{1}{4}\bigl[\sin^{2}\theta\left\langle \sigma_{i}^{y}\sigma_{j}^{y}\right\rangle -\sin\theta\cos\theta\left\langle \sigma_{i}^{y}\sigma_{j}^{z}+\sigma_{i}^{z}\sigma_{j}^{y}\right\rangle
+\cos^{2}\theta\left\langle \sigma_{i}^{z}\sigma_{j}^{z}\right\rangle \bigr].
\end{equation}
These so-called spin-squeezing measurements \cite{Uys_etal} are sensitive to pairwise correlations summed over all the spins in the ensemble.

\section{Outlook and summary.}
Once such a quantum simulator is benchmarked, it may in turn be used for the testing of approximate calculations of the quantum dynamics of the long-range Ising model in the presence of non-longitudinal fields (resulting in non-commuting terms in the Hamiltonian \eref{e:H1}). Moreover, there are certain phenomena which are believed to be peculiar to long-range interacting systems and which might receive their first experimental confirmation in the ion trap setup. One example is the threshold at $\alpha=d/2$ below which a second timescale of relaxation emerges. There is evidence that not only the existence of this dynamical long-range threshold, but also its numerical value $d/2$, is universal for classical as well as quantum mechanical long-range interacting systems \cite{BachelardKastner13}. Other long-range peculiarities include nonequivalent equilibrium statistical ensembles \cite{Kastner10,KastnerJSTAT10}, a phenomenon that has been known in the astrophysical context for decades \cite{Thirring70} but has not seen experimental verification in a tabletop experiment.  

In summary, under convenient restrictions on the initial conditions, we have obtained exact analytic results for spin--spin correlations of long-range quantum Ising models in arbitrary spatial dimensions and for various lattice structures. Pre\-ther\-mal\-i\-za\-tion, widely separated timescales, and other phenomena that further our understanding of the timescales governing the relaxation to equilibrium were discussed. Comparison of our analytic results to experimental measurements of spin--spin correlations will enable benchmarking of the trapped-ion quantum simulator of \cite{Britton_etal12} in a regime where quantum effects are paramount. Subsequently, the benchmarked quantum simulator may be employed to investigate relaxation timescales in parameter regimes and for initial states where analytic results presently are not available.

\appendix

\section{Exact calculation of spin--spin correlations for finite \texorpdfstring{$N$}{N}.}
\label{A}
Starting point of the calculation is the general expression of the expectation value
\begin{equation}\label{e:spinspin}
\left\langle\sigma_i^\pm\sigma_j^\pm\right\rangle\!(t)=\Tr \left(\rme^{\rmi H_\ell t} \sigma_i^\pm\sigma_j^\pm \rme^{-\rmi H_\ell t}\rho_0\right)
\end{equation}
with respect to the initial density operator $\rho_0$, where we assume $\rho_0$ to be diagonal in the $\sigma_x$ tensor product eigenbasis. Spin ladder operators are defined as $\sigma^\pm=(\sigma^x\pm\rmi\sigma^y)/2$. Since all terms in the Hamiltonian $H_\ell$ commute, we can factorize the time evolution operator,
\begin{equation}\label{e:U}
\exp\left(-\rmi H_\ell t\right)=\prod_{k<l}\exp\left(\rmi J_{k,l}\sigma_k^z\sigma_l^z t\right)\prod_m\exp\left(\rmi B\sigma_m^z t\right),
\end{equation}
and similarly for the Hermitian conjugate. All factors in \eref{e:U} that contain neither $\sigma_i^z$ nor $\sigma_j^z$ commute with $\sigma_i^\pm\sigma_j^\pm$. To compute the time evolution in \eref{e:spinspin}, we therefore have to deal with the expression
\begin{eqnarray}\label{e:ssoft}
\fl\prod_{k\neq i,j}\exp\left(-\rmi J_{k,i}\sigma_k^z\sigma_i^z t\right)\exp\left(-\rmi J_{k,j}\sigma_k^z\sigma_j^z t\right)\exp\left(-\rmi B\sigma_i^z t\right)\exp\left(-\rmi B\sigma_j^z t\right)\sigma_i^\pm\sigma_j^\pm\nonumber\\
\times\exp\left(\rmi B\sigma_j^z t\right)\exp\left(\rmi B\sigma_i^z t\right)\prod_{l\neq i,j}\exp\left(\rmi J_{l,j}\sigma_l^z\sigma_j^z t\right)\exp\left(\rmi J_{l,i}\sigma_l^z\sigma_i^z t\right).
\end{eqnarray}
Making use of $\left[\sigma^z,\sigma^\pm\right]=\pm2\sigma^\pm$, the time evolution due to the magnetic field $B$ simplifies to
\begin{equation}\label{e:Bevolv}
\fl\exp\left(-\rmi B\sigma_i^z t\right)\exp\left(-\rmi B\sigma_j^z t\right)\sigma_i^\pm\sigma_j^\pm\exp\left(\rmi B\sigma_j^z t\right)\exp\left(\rmi B\sigma_i^z t\right)=\sigma_i^\pm\sigma_j^\pm\exp\left(\mp4\rmi Bt\right).
\end{equation}
Picking one lattice site $k\neq i,j$, the time evolution of $\sigma_i^\pm\sigma_j^\pm$ due to the interaction with the spin at $k$ can be written as
\begin{eqnarray}\label{e:Jevolv}
\fl\exp\left(-\rmi J_{k,i}\sigma_k^z\sigma_i^z t\right)\exp\left(-\rmi J_{k,j}\sigma_k^z\sigma_j^z t\right)\sigma_i^\pm\sigma_j^\pm
\exp\left(\rmi J_{k,j}\sigma_k^z\sigma_j^z t\right)\exp\left(\rmi J_{k,i}\sigma_k^z\sigma_i^z t\right)\nonumber\\
=\sigma_i^\pm\sigma_j^\pm\cos\left[2t(J_{i,k}+J_{j,k})\right]\mp \rmi\sigma_i^\pm\sigma_j^\pm\sigma_k^z\sin\left[2t(J_{i,k}+J_{j,k})\right].
\end{eqnarray}
Since the initial state $\rho_0$ is assumed to be diagonal in the $\sigma_x$ tensor product eigenbasis, only diagonal elements of the operator $\rme^{\rmi H_\ell t} \sigma_i^\pm\sigma_j^\pm \rme^{-\rmi H_\ell t}$ in the same basis contribute to the trace in \eref{e:spinspin}. For this reason we can drop the second term on the right-hand side of \eref{e:Jevolv}, as it is proportional to $\sigma_k^z$. Inserting \eref{e:Bevolv} and \eref{e:Jevolv} into \eref{e:spinspin}, we obtain
\begin{equation}\label{e:pmpmfinal}
\left\langle\sigma_i^\pm\sigma_j^\pm\right\rangle\!(t)=\left\langle\sigma_i^\pm\sigma_j^\pm\right\rangle\!(0)\exp\left(\mp4\rmi Bt\right)
\prod_{k\neq i,j}\cos\left[2t(J_{i,k}+J_{j,k})\right],
\end{equation}
where $\left\langle\sigma_i^\pm\sigma_j^\pm\right\rangle\!(0)=\Tr\left(\sigma_i^\pm\sigma_j^\pm\rho_0\right)$. A similar calculation yields
\begin{equation}\label{e:pmmpfinal}
\left\langle\sigma_i^\pm\sigma_j^\mp\right\rangle\!(t)=\left\langle\sigma_i^\pm\sigma_j^\mp\right\rangle\!(0)\prod_{k\neq i,j}\cos\left[2t(J_{i,k}- J_{j,k})\right],
\end{equation}
a notable difference to \eref{e:pmpmfinal} being that the presence of an external magnetic field $B$ has no effect on this expectation value. From
\begin{eqnarray}
4\sigma_i^\pm\sigma_j^\pm&=&\sigma_i^x\sigma_j^x-\sigma_i^y\sigma_j^y\pm\rmi\sigma_i^x\sigma_j^y\pm\rmi\sigma_i^y\sigma_j^x,\\
4\sigma_i^\pm\sigma_j^\mp&=&\sigma_i^x\sigma_j^x+\sigma_i^y\sigma_j^y\mp\rmi\sigma_i^x\sigma_j^y\pm\rmi\sigma_i^y\sigma_j^x,
\end{eqnarray}
we obtain
\begin{eqnarray}
\sigma_i^x\sigma_j^x&=&\phantom{\mp}2\Re\left(\sigma_i^\pm\sigma_j^\mp+\sigma_i^\pm\sigma_j^\pm\right),\label{e:xx}\\
\sigma_i^y\sigma_j^y&=&\phantom{\mp}2\Re\left(\sigma_i^\pm\sigma_j^\mp-\sigma_i^\pm\sigma_j^\pm\right),\label{e:yy}\\
\sigma_i^x\sigma_j^y&=&\mp2\Im\left(\sigma_i^\pm\sigma_j^\mp-\sigma_i^\pm\sigma_j^\pm\right),\label{e:xy}\\
\sigma_i^y\sigma_j^x&=&\pm2\Im\left(\sigma_i^\pm\sigma_j^\mp+\sigma_i^\pm\sigma_j^\pm\right),\label{e:yx}
\end{eqnarray}
and furthermore, using the properties of the initial density operator $\rho_0$,
\begin{equation}\label{e:initial}
4\left\langle\sigma_i^\pm\sigma_j^\pm\right\rangle\!(0) = 4\left\langle\sigma_i^\pm\sigma_j^\mp\right\rangle\!(0) = \left\langle\sigma_i^x\sigma_j^x\right\rangle\!(0).
\end{equation}
Inserting \eref{e:pmpmfinal}, \eref{e:pmmpfinal} and \eref{e:initial} into \eref{e:xx}--\eref{e:yx}, we arrive at the final expressions
\begin{eqnarray}
\left\langle\sigma_i^x\sigma_j^x\right\rangle\!(t)&=&P_{i,j}^- + \cos(4Bt)P_{i,j}^+,\\
\left\langle\sigma_i^y\sigma_j^y\right\rangle\!(t)&=&P_{i,j}^- - \cos(4Bt)P_{i,j}^+,\\
\left\langle\sigma_i^x\sigma_j^y\right\rangle\!(t)&=& -\sin(4Bt)P_{i,j}^+,
\end{eqnarray}
with $P_{i,j}^\pm$ as defined in \eref{e:Ppm}. A similar calculation yields
\begin{equation}
\left\langle\sigma_i^x\sigma_j^z\right\rangle\!(t)=\sin(2Bt)P_{i,j}^z,\qquad \left\langle\sigma_i^y\sigma_j^z\right\rangle\!(t)=\cos(2Bt)P_{i,j}^z,
\end{equation}
with
\begin{equation}
P_{i,j}^z=-\left\langle\sigma_i^x\right\rangle\!(0)\sin(2tJ_{i,j})\prod_{k\neq i,j}\cos(2tJ_{i,k}).
\end{equation}
Finally, from symmetry considerations we obtain
\begin{equation}
\left\langle\sigma_i^z\sigma_j^z\right\rangle\!(t)=0.
\end{equation}

A generalization of these calculations to $n$-spin correlation functions is straight\-forward, yielding for example
\begin{equation}\label{e:pmpmpm}
\fl\left\langle\sigma_i^\pm\sigma_j^\pm\sigma_k^\pm\right\rangle\!(t)=\left\langle\sigma_i^\pm\sigma_j^\pm\sigma_k^\pm\right\rangle\!(0)\exp\left(\mp6\rmi Bt\right)
\prod_{l\neq i,j,k}\cos\left[2t(J_{i,l}+J_{j,l}+J_{k,l})\right],
\end{equation}
and similar expressions for other correlation functions.

\section{Upper bounds on correlations in the thermodynamic limit.}
\label{B}
In the thermodynamic limit of infinite system size, a bound on the product
\begin{equation}\label{e:Ppm_def}
\mathcal{P}_{i,j}^\pm=\prod_{\scriptstyle k=1\atop \scriptstyle k\neq i,j}^N\left|\cos\left[2\left(J_{k,i}\pm J_{k,j}\right)t\right]\right|
\end{equation}
in equation \eref{e:Ppm} 
with $J_{i,j}=D_{i,j}^{-\alpha}$ and $\alpha\geqslant0$ is derived. For any given $t$, one can find a compact region $g_{i,j}^\pm(t)$ (containing $i$ and $j$) of the infinite triangular lattice $G$ such that
\begin{equation}\label{e:JJ}
|2(J_{k,i}\pm J_{k,j})t|<\pi/2
\end{equation}
for all $k\in G\setminus g_{i,j}^\pm(t)$. Since $|\cos x|\leqslant1$, we have 
\begin{equation}
\mathcal{P}_{i,j}^\pm\leqslant\prod_{k\notin g_{i,j}^\pm(t)}\left|\cos\left[2\left(J_{k,i}\pm J_{k,j}\right)t\right]\right|.
\end{equation}
Next, using the inequality
\begin{equation}
|\cos x|\leqslant1-\left(\frac{2x}{\pi}\right)^2\leqslant\exp\biggl[-\left(\frac{2x}{\pi}\right)^2\biggr]\qquad\mbox{for $x<\pi/2$},
\end{equation}
we obtain
\begin{equation}\label{e:expbound}
\mathcal{P}_{i,j}^\pm\leqslant\exp\Biggl[-\frac{16t^2}{\pi^2}\sum_{k\notin g_{i,j}^\pm(t)}\left(J_{k,i}\pm J_{k,j}\right)^2\Biggr].
\end{equation}

For simplicity, we consider lattice sites $i$ and $j$ symmetrically arranged to the right and left of lattice site 0, as illustrated below for the example of distance $\delta=D_{i,j}=2$, but other arrangements can be treated in a similar way. 
\begin{center}
\includegraphics[width=0.65\linewidth]{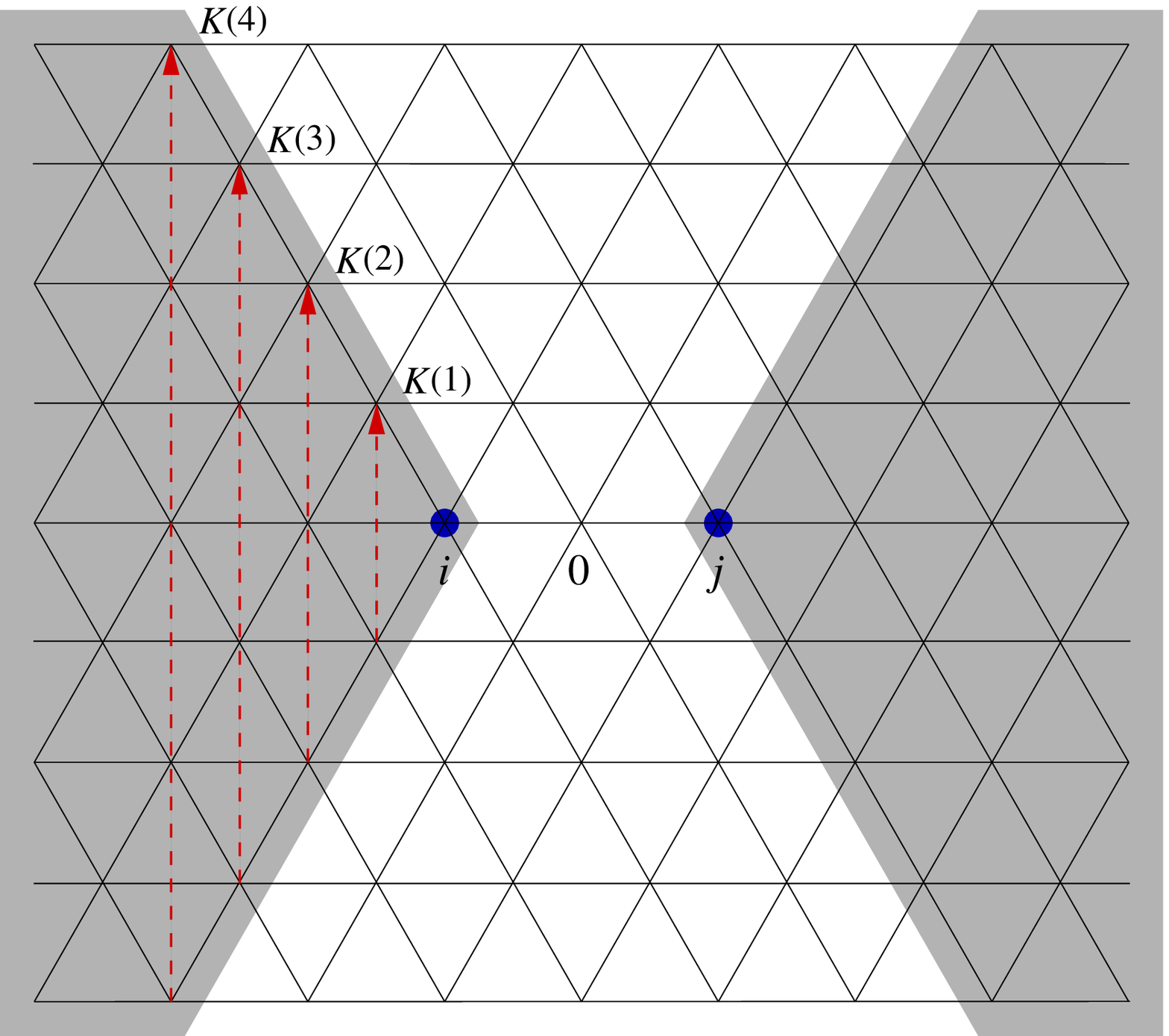}
\end{center}
Disregarding all sites that lie outside the grey shaded area $h_{i,j}$ in the above illustration, we obtain a lower bound on the sum in \eref{e:expbound},
\begin{equation}
\sum_{k\in G\setminus g_{i,j}^\pm(t)}\left(J_{k,i}\pm J_{k,j}\right)^2\geqslant\sum_{k\in h_{i,j} \cap (G\setminus g^{\pm}_{i,j}(t))}\left(J_{k,i}\pm J_{k,j}\right)^2.
\end{equation}
Each individual summand $\left(J_{k,i}\pm J_{k,j}\right)^2$ can be minorized by $\left(J_{K,i}\pm J_{K,j}\right)^2$, where $K$ is determined from $k$ by going vertically upwards along the dashed (red) line in the above illustration until a boundary point of the shaded region is reached. To a boundary point labelled by $K(r)$ there correspond $r+1$ points in the sum, and this allows us to bound the sum by
\begin{equation}
\sum_{k\in G\setminus g_{i,j}^\pm(t)}\left(J_{k,i}\pm J_{k,j}\right)^2\geqslant2J^2\sum_{r=R_0^\pm(t)}^R r\left(D_{i,K(r)}^{-\alpha}\pm D_{j,K(r)}^{-\alpha}\right)^2\!\!.
\end{equation}
\begin{figure}
{\center
\includegraphics[width=0.48\linewidth]{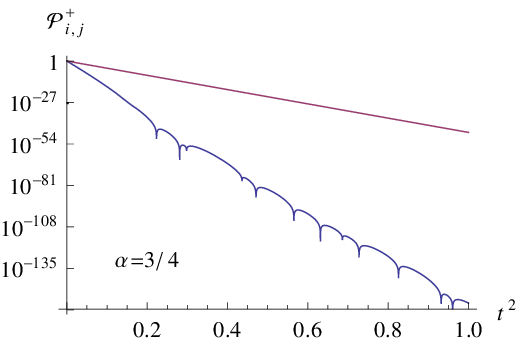}
\hfill
\includegraphics[width=0.48\linewidth]{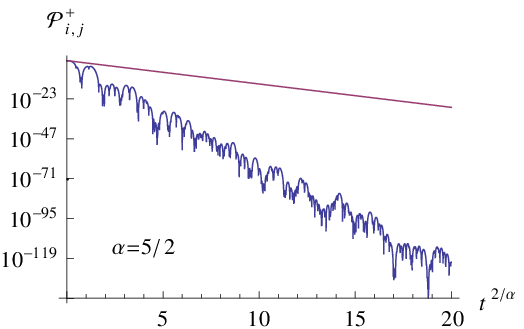}\\
\includegraphics[width=0.48\linewidth]{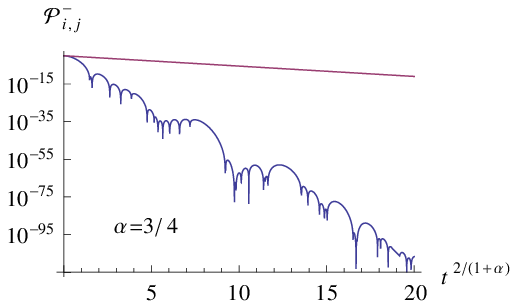}
\hfill
\includegraphics[width=0.48\linewidth]{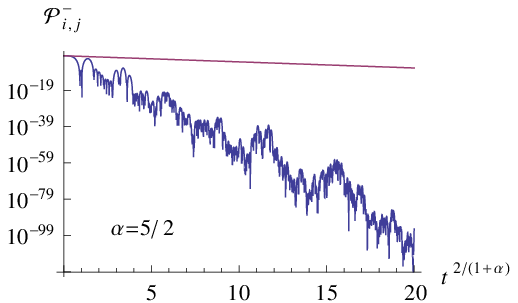}
}
\caption{\label{f:bounds}%
Logarithmic plot of the products $\mathcal{P}_{i,j}^+$ (top) and $\mathcal{P}_{i,j}^-$ (bottom) as a function of rescaled time $t^{p^\pm(\alpha)}$, evaluated for a hexagonal patch of a triangular lattice with side length $L=16$ for $\alpha=3/4$ (left) and $\alpha=5/2$ (right). The exponents $p^+(\alpha)=\min\{2,2/\alpha\}$ and $p^-(\alpha)=2/(1+\alpha)$ of the time rescaling are those predicted by the bounds \eref{e:expboundfinal+} and \eref{e:expboundfinal-} for the asymptotic behaviour at large $L$ and $t$. The linearly decaying trend in the plot, superimposed by fluctuations, confirms over a range of more than hundred orders of magnitude that the asymptotic behaviour of $\mathcal{P}_{i,j}^\pm$ is indeed as given by the bounds. The straight lines in the plots are the bounds \eref{e:expboundfinal+} and \eref{e:expboundfinal-}, indicating that the numerical constants in the exponents of \eref{e:expboundfinal+} and \eref{e:expboundfinal-} are, as expected, underestimated.
}%
\end{figure}
To exclude lattice sites in the region $g_{i,j}^\pm(t)$, i.e.\ in order to satisfy the inequality \eref{e:JJ} used earlier, $R_0^\pm(t)$ has to be chosen large enough. An asymptotic analysis of \eref{e:JJ} shows that this condition can be implemented by choosing
\begin{equation}
R_0^+(t)>\left(\frac{8Jt}{\pi}\right)^{1/\alpha},\qquad R_0^-(t)\sim\left(\frac{4\alpha Jt}{\pi}\right)^{1/(1+\alpha)},
\end{equation}
where the condition on $R_0^+(t)$ works always, the one for $R_0^-(t)$ only for sufficiently large $t$. Inserting the distances
\begin{equation}
D_{i,K(r)}=r,\qquad D_{j,K(r)}=\sqrt{\delta^2+r\delta+r^2}
\end{equation}
and bounding the sum by an integral, we obtain
\begin{equation}\label{e:PIntegral}
\mathcal{P}_{i,j}^\pm\leqslant\exp\Biggl[-\frac{32J^2t^2}{\pi^2}\int\limits_{R_0^\pm(t)}^R\rmd r\, r \left(\frac{1}{r^\alpha}\pm\frac{1}{\sqrt{\delta^2+r\delta+r^2}^\alpha}\right)^2\Biggr],
\end{equation}
valid in the limit of large $R$. Since we are interested in the limit of large $R$ and $t$, we can expand the integrand to leading order in $1/r$, obtaining
\begin{eqnarray}
r \left(\frac{1}{r^\alpha}+\frac{1}{\sqrt{\delta^2+r\delta+r^2}^\alpha}\right)^2 & \sim & 4r^{1-2\alpha},\\
r \left(\frac{1}{r^\alpha}-\frac{1}{\sqrt{\delta^2+r\delta+r^2}^\alpha}\right)^2 & \sim & \left(\frac{\alpha\delta}{2}\right)^2 r^{-1-2\alpha}.
\end{eqnarray}
Inserting these expressions into \eref{e:PIntegral}, the integral can be solved by elementary means. Making use of the asymptotic equality $R^2\sim N/3$ and performing the limit $N\to\infty$, we obtain the final results
\begin{equation}\label{e:expboundfinal+}
\mathcal{P}_{i,j}^+\leqslant
\cases{\exp\left[-\frac{64J^2t^2}{\pi^2(1-\alpha)}\left(\frac{N}{3}\right)^{1-\alpha}\right] & for $\alpha<1$,\\
\displaystyle\exp\left[-\frac{1}{\alpha-1}\left(\frac{8Jt}{\pi}\right)^{2/\alpha}\right] & for $\alpha>1$,\\}
\end{equation}
and
\begin{equation}\label{e:expboundfinal-}
\mathcal{P}_{i,j}^-\leqslant\exp\left[-\frac{\delta^2}{4\alpha}\left(\frac{4\alpha Jt}{\pi}\right)^{2/(1+\alpha)}\right],
\end{equation}
valid in the limit of large $R$ and $t$. A comparison of these bounds with an exact evaluation of \eref{e:Ppm_def} for finite lattices is shown, over more than hundred orders of magnitude, in \fref{f:bounds}.


\ack
\addcontentsline{toc}{section}{Acknowledgments}
BCS and JJB acknowledge support from the DARPA OLE program. MK acknowledges support by the Incentive Funding for Rated Researchers programme of the National Research Foundation of South Africa. Manuscripts with contributions from the US National Institute of Standards and Technology are not subject to US copyright. During the preparation of this manuscript we were informed of related work on spin--spin correlation functions of long-range Ising models \cite{Foss-Feig_etal13}. In contrast to our work, which focuses on large time asymptotic behaviour and relaxation to equilibrium, Foss-Feig {\em et al}\/ \cite{Foss-Feig_etal13} investigate the effect of decoherence on the spin--spin correlation functions.\\


\addcontentsline{toc}{section}{References}
\bibliographystyle{iopart-num}
\bibliography{Correlators}

\end{document}